\documentclass[lettersize,journal]{IEEEtran}
\usepackage{lineno,hyperref}
\usepackage{float}
\usepackage{graphicx}
\usepackage{epstopdf}
\usepackage{algorithm}
\usepackage{algorithmicx}
\usepackage{algpseudocode}
\usepackage{amsmath}
\usepackage{amstext}
\usepackage{amssymb}
\usepackage{caption}
\usepackage{subfigure}
\usepackage{color}
\usepackage{multirow}
\usepackage{subfigure}
\usepackage{listings}
\usepackage{xcolor}
\usepackage[numbers,sort&compress]{natbib}
\usepackage{setspace}
\usepackage{makecell}
\usepackage[english]{babel}
\begin{document}

\title{A Blockchain-based Semantic Exchange Framework for Web 3.0 toward Participatory Economy}

\author{
  Yijing Lin,
  Zhipeng Gao*,
  Yaofeng Tu,
  Hongyang Du,
  Dusit Niyato,
  Jiawen Kang,
  Hui Yang
}
\maketitle

\begin{abstract}

Web 3.0 is the next-generation Internet that enables participants to read, write, and own contents in a decentralized manner. It is mainly driven by blockchain, semantic communication, edge computing, and artificial intelligence, which can construct value networks to achieve participatory economics based on participatory decision making. Web 3.0 can capture the characteristics of blockchain, semantic extraction, and communication to achieve decentralized semantic sharing and transfer information precisely. However, current Web 3.0 solutions focus on the blockchain while overlooking other new technologies' roles in Web 3.0. To further unleash the advantages of semantic extraction and communication in Web 3.0, in this paper, we propose a blockchain-based semantic exchange framework to realize fair and efficient interactions. In this framework, we first attempt to tokenize semantic information into Non-Fungible Token (NFT) for semantic exchange. Then we utilize a Stackelberg game to maximize buying and pricing strategies for semantic trading. We also leverage Zero-Knowledge Proof to share authentic semantic information without publishing it before receiving payments, which can achieve a fair and privacy-preserving trading compared with current NFT marketplaces. A case study about urban planning is given to show clearly the proposed mechanisms. Finally, several challenges and opportunities are identified.
\end{abstract}


\section{Introduction}
\label{sec_introduction}

Web 3.0 is now attracting the interest of researchers from academics and industries. It is considered to be the next generation of the Internet where participants can control their data, identities, and capabilities to make decentralized decisions and construct participatory economy. There are two mainstream developments of Web 3.0 as follows, including Blockchain Web 3.0 and Semantic Web 3.0. When the concept of Web 3.0 first emerged, it was identified as a semantic web, which can make Internet data machine-readable \cite{berners2001semantic} and reduce energy consumption for processing and transmission. The current definition of Web 3.0 \cite{wood_2014} was prevailing with the help of blockchain technologies, which can make participants get rid of tech giants and control user-generated content. Besides, since artificial intelligence and edge computing play important roles to connect Web 3.0 participants in a smart and lightweight way, they are also considered vital components of Web 3.0. 

The diversified underlying technologies of Web 3.0 are similar to Web 2.0 mainly driven by cloud computing, social networks, and mobile connectivity. Instead of being controlled by tech giants, Web 3.0 allows all participants to make decisions to participate in constructing value networks, which should permit participants without the need of having the expertise of underlying technologies, like DeFi Money Legos \cite{cousaert2022sok}. Thus, Web 3.0 can be supported by multiple technologies like blockchain, artificial intelligence, edge computing, and semantic communication, which can easily coordinate to construct value networks to achieve participatory economy based on decentralized decision making among participants. 

Although several new technologies are emerging as main components to construct value networks in Web 3.0, blockchain is considered the main technology for the participatory economy in many cases. Few researchers focus on how to integrate the components of new technologies into blockchain-based Web 3.0 toward the participatory economy. DeCAST \cite{zhao2022decast} utilizes the integration of blockchain and federated learning to construct parallel intelligent transportation systems, which are effective and practical for smart mobility. A unified blockchain-semantic framework \cite{lin2022unified} is proposed to capture the benefits of blockchain, semantic extraction, and communication to improve interaction efficiency for Web 3.0. Besides, given the steps of the integration of new technologies in Web 3.0, it is necessary to consider how to make value circulated in the participatory economy to attract participants to join in Web 3.0 decision-making.

\begin{figure}[!t]
  \centering
  \includegraphics[height=3in]{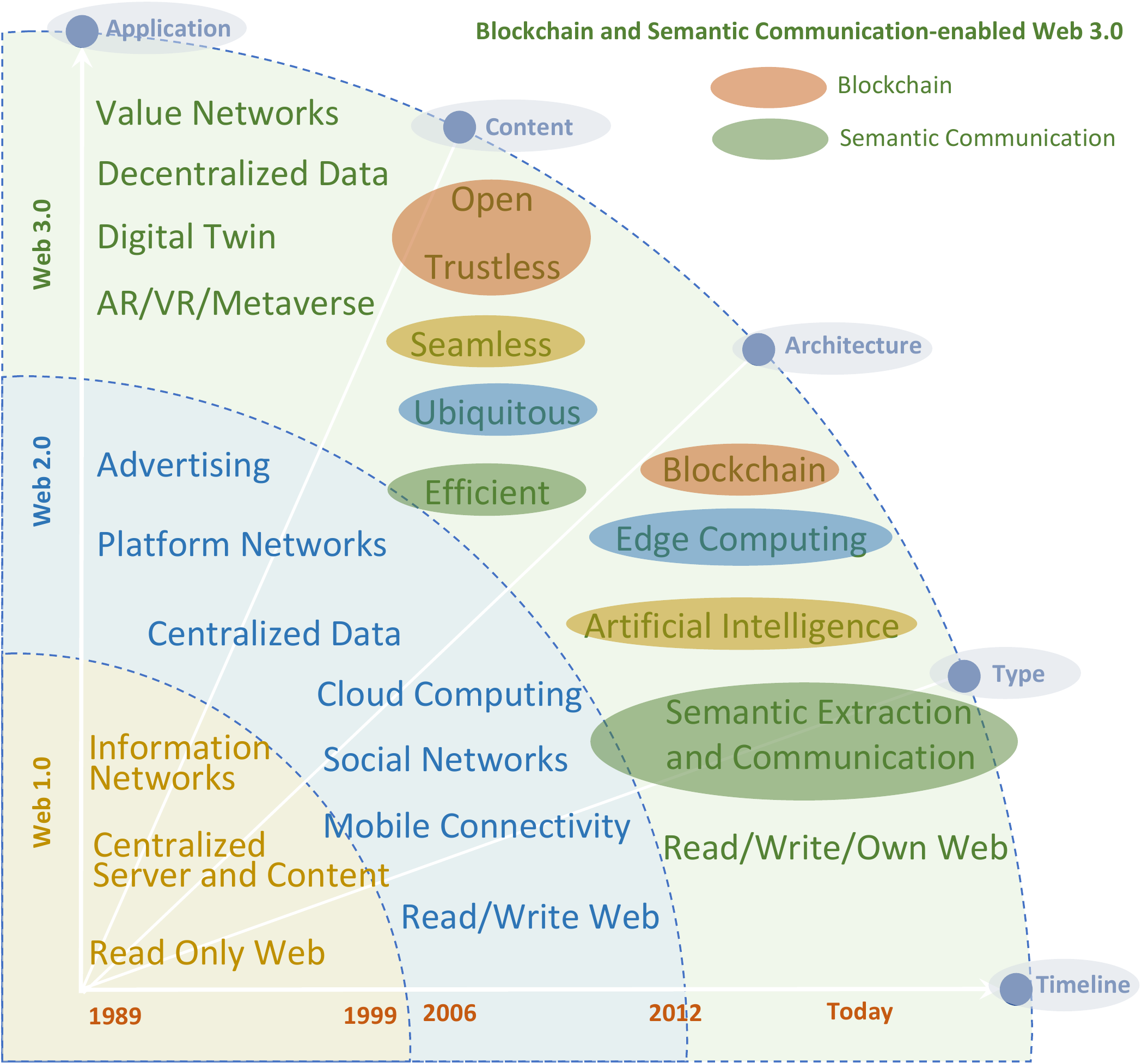}
  \caption{Difference Between Web 1.0, Web 2.0 and Web 3.0}
  \label{fig_diff}
\end{figure}
\begin{figure*}[!t]
  \centering
  \includegraphics[height=3.5in,width=7in]{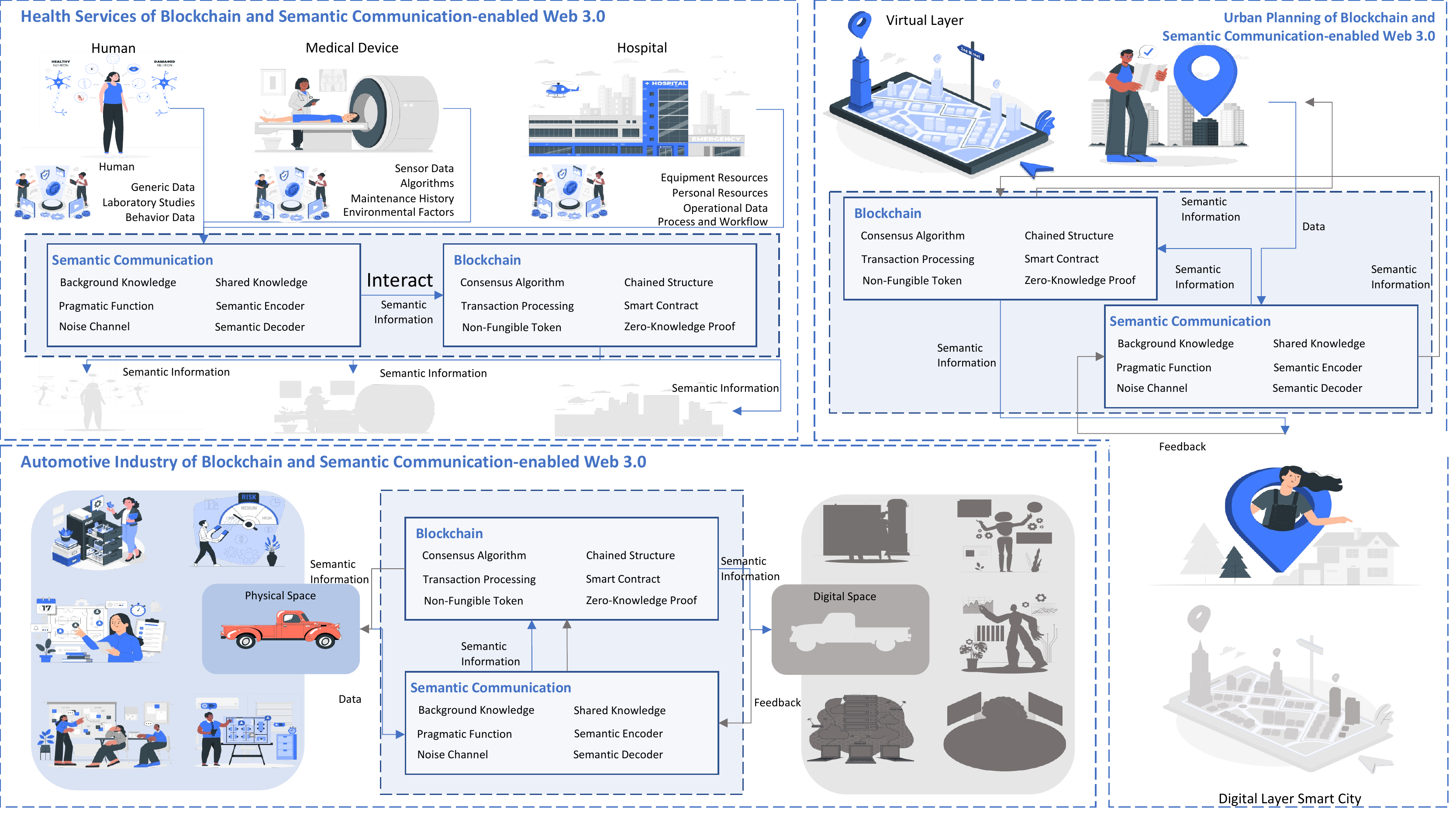}
  \caption{Applications of Blockchain and Semantic Communication-enabled Web 3.0}
  \label{fig_digital_twin}
\end{figure*}

Blockchain, semantic extraction and communication can be utilized to reduce information overloaded, reduce energy consumption, and construct efficient interactions toward the participatory economy by extracting semantic information from raw data and implementing sharing in a decentralized manner. However, even capitalizing on the benefits of the above technologies, the design toward the participatory economy still faces many difficulties. On the one hand, it is difficult to quantify and transact semantic information circulated in Web 3.0. The reason is that producers and consumers are difficult to maximize their revenue and utility \cite{li2022multi}. On the other hand, Web 3.0 is supported by a public blockchain platform, which makes pending semantic information exposed to both parties. Consumers can copy semantic information without payments to maximize their benefits \cite{song2022traceable} in current marketplaces.

In this paper, we attempt to complement the understanding of blockchain, semantic extraction and communication playing an important role in Web 3.0. We also develop a fair and efficient mechanism based on blockchain and semantic communication to tokenize semantic information and enable semantic exchange by utilizing the Non-Fungible Token (NFT), Stackelberg Game, and Zero Knowledge Proof (ZKP) toward participatory economy. Our contributions are summarized as follows.

\begin{itemize}
	\item We propose a new semantic exchange framework, that tokenize semantic information to facilitate semantic exchange by using blockchain.	
	\item In this framework, we formulate a Stackelberg game-based semantic trading mechanism to design optimal pricing strategies.
	\item To facilitate fair and privacy-preserving transactions between producers and consumers, we utilize zero-knowledge proof to transact transformed semantic information without publishing authentic one before payments.
	\item To show clearly the applications and mechanisms, we utilize urban planning to display how the above contributions work.
\end{itemize}

\section{The Future of Blockchain, Semantic Extraction and Communication-enabled the Internet}
\label{related_work}

\subsection{An Accelerating Growth of The Internet}

The accelerating growth of the iterations of the Internet is witnessed from Web 1.0 to Web 3.0, as shown in Fig. \ref{fig_diff}. According to Fig. \ref{fig_diff}, we illustrate the difference between Web 1.0, Web 2.0, and Web 3.0, the connection between Web 3.0, blockchain, edge computing, and semantic communication, and the applications of blockchain and semantic communication-enabled Web 3.0. 

\subsubsection{What Is The Difference Between Web 1.0, Web 2.0, And Web 3.0} 

The development of the Internet experiences three iterations of Web 1.0, Web 2.0, and Web 3.0 eras. Web 1.0 is a read-only web, where users can only read information produced by centralized content producers from centralized servers. Web 2.0 is a read-write web driven by mobile connectivity, social networks, and cloud computing, where users can read and write content on platforms provided by giant tech companies. Platforms can utilize content produced by users to make a profit like advertising without paying any to producers. Therefore, the need for Web 3.0 is that users expect to control their data and protect their privacy in a decentralized manner. Web 3.0 is a read-write-own web driven by blockchain, edge computing, artificial intelligence, and semantic communication, where users can read, write, and own content on decentralized value networks.

\subsubsection{What Is The Connection Between Web 3.0, Blockchain, Edge Computing, And Semantic Communication} 

Whereas Web 2.0 was mainly driven by mobile connectivity, social networks, and cloud computing to allow users to participate in constructing platform networks, Web 3.0 is enabled by blockchain, edge computing, artificial intelligence, and semantic communication to design an open, trustless, seamless, ubiquitous, and efficient next-generation Internet. Blockchain is often considered the same as Web 3.0, while it is actually an infrastructure to support decentralized data sharing in an open and trustless manner. Edge computing is also vital to construct Web 3.0, which can leverage the computing and storage capabilities of ubiquitous edge devices and servers to make Internet services available and near to users. Artificial intelligence can leverage tons of data produced by edge computing and blockchain to train models, which can improve services and provide seamless experiences. Instead of transmitting information accurately, semantic communication \cite{yang2022semantic} based on edge computing and artificial intelligence pursues how precisely the transmitted information can convey the desired meanings to reduce information overloaded and energy consumed to provide Web 3.0 with efficiency. 

\subsubsection{What Are Applications of Blockchain And Semantic Communication-enabled Web 3.0} Whereas Web 2.0 mainly focuses on applications on platform networks for controllers to make a profit from user data, blockchain and semantic communication-enabled Web 3.0 concentrate on value networks to facilitate transparent use of data, and decentralized data exchange and sharing. Blockchain and semantic communication-enabled Web 3.0 toward participatory economy can be applied for embracing the digital transformation of healthcare services, the automotive industry, and urban planning, as shown in Fig. \ref{fig_digital_twin}.



\subsection{Research Gaps of Blockchain and Semantic Communication-enabled Web 3.0}

Instead of a mansion in the air, Web 3.0 is the next-generation infrastructure consisting of multiple building blocks like blockchain and semantic communication, which is also similar to blockchain made up of cryptography and distributed algorithms. Web 3.0 is expected to construct an entirely new Internet for participants to transact and interact in the participatory economy.

Previous works only focus on the blockchain or semantic-enabled Web 3.0, while there are few works focusing on the integration between Web 3.0, blockchain, semantic extraction and communication. DeSci \cite{ding2022desci} utilized blockchain and decentralized autonomous organizations to construct a reference model for scientific systems. A unified blockchain-semantic framework enabled Web 3.0 was proposed to maintain service security and improve interaction efficiency in Web 3.0 \cite{lin2022unified}. HyperService \cite{liu2021make} was designed to construct an interoperable cross-chain platform for decentralized applications to interact data across blockchains.

Besides, blockchain and semantic-enabled Web 3.0 still suffers from the following challenges: 1) how to release the value of semantic information when combining semantic communication and blockchain with Web 3.0, 2) how to maximize the revenue and utility of semantic information producers and consumers in the participatory economy, 3) how to trade semantic information among producers and consumers in a fair manner. To solve the above challenges, we propose a fair and efficient blockchain-based semantic exchange framework for Web 3.0 by NFT-based semantic exchange, Stackelberg game-based semantic trading, and zero-knowledge proof-based semantic sharing mechanisms to form the participatory economy.

\section{Fair and Efficient Mechanism Based on Semantic Blockchain}
\label{sec_semantic_blockchain}

In this section, we describe the NFT-based semantic exchange scheme to facilitate semantic circulation. We also propose a Stackelberg game-based dynamic semantic trading mechanism to maximize the utility and revenue of semantic information. Moreover, we construct a zero-knowledge proof-based semantic sharing mechanism to enable fair authentic semantic sharing. 

\subsection{NFT-based Semantic Exchange Scheme}

Semantic information extracted from producers and exchanged in semantic communications has been considered how to determine the importance of information \cite{yang2022semantic}. Thus, for semantic information of high importance, it is necessary to decide digital ownerships to facilitate the circulation of valuable semantic information. Therefore, we propose an NFT-based semantic exchange scheme to tokenize semantic information and unalterably prove the digital ownership of producers.


  \begin{figure*}[!t]
  \centering
  \includegraphics[height=3.5in,width=7in]{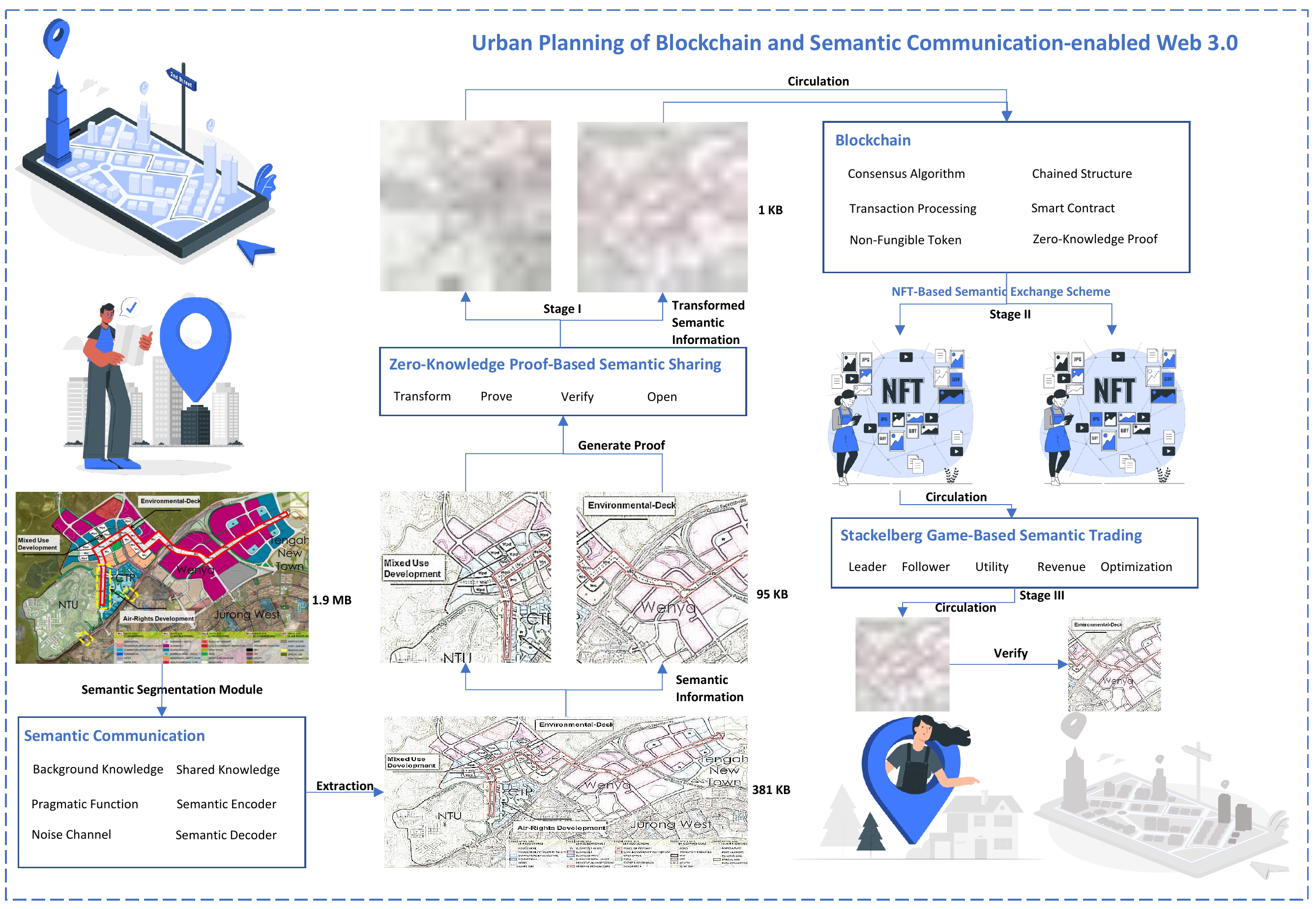}
  \caption{Case Study in Urban Planning}
  \label{fig_case_study}
\end{figure*}

The NFT-based semantic exchange scheme is composed of three functional parts, including blockchain, off-chain storage, and semantic communication systems. The blockchain is constructed by edge servers to enable consensus and smart contracts, which can support NFT-based semantic exchange transactions between edge devices. Since blockchain has limited storage spaces, off-chain storage is utilized to supplement spaces for blockchain to interact with texts, images, and videos in an efficient way. Edge devices can contribute spare storage spaces to extend the off-chain storage to expand the storage ability of blockchain. Once semantic information is uploaded into off-chain storage, the off-chain storage will return a hash value to uniquely map the semantic information to prevent tampering. The semantic communication systems include semantic information producers and consumers to extract, exchange, and consume semantic information to reduce information overloaded. Moreover, since semantic communication systems need to interact with the blockchain and off-chain storage, edge devices (producers and consumers) should equip with key generation, blockchain client, and off-chain storage client modules to implement semantic information circulating. The workflow of the NFT-based semantic exchange scheme is described as follows.

\begin{itemize}
  \item \textbf{Step 1: Write and Deploy NFT Smart Contracts.} NFTs are on-chain credentials for semantic information and proofs of the ownership of producers, which are implemented by smart contracts. The function of NFT smart contracts mainly include \textit{Mint}, \textit{Transfer}, and \textit{Burn}. 
  
  \textit{Mint} enables producers to utilize smart contracts to release NFT tokens mapping semantic information which is stored in off-chain storage. \textit{Transfer} can shift the ownership of NFT tokens from producers to consumers, which means the exchange of semantic information. \textit{Burn} destroys NFT tokens on the blockchain, which excludes semantic information from circulation between producers and consumers.

  \item \textbf{Step 2: Publish Semantic information.} Due to the limited storage of blockchain, semantic information is first published in off-chain storage by producers via IPFS clients. Producers obtain metadata to access the semantic information via Distributed Hash Table. Metadata can be considered as the hash commitment of semantic information to prevent it from tampering.
  
  \item \textbf{Step 3: Mint Semantic information into NFT.} Producers can bind metadata of semantic information, mint them into NFT tokens, and generate token URIs by smart contracts and blockchain clients. Besides metadata, NFT tokens should contain attributes to introduce types of semantic tasks, producers, and time, which can be transmitted in JSON format.  
  
  \item \textbf{Step 4: Transfer NFT-mapped Semantic Information.} Producers can send a \textit{transferFrom} transaction to NFT smart contract to assign NFT tokens to consumers, which can exchange semantic information stored in off-chain storage. When consumers own NFT tokens mapped semantic information, they can transfer it to other consumers to make a profit. NFT tokens can be observed to track how the ownership of semantic information changes over time.
\end{itemize}

Besides, producers can send approved transactions to authorize trading marketplaces as operators, and operators can substitute for owners to transfer semantic information from producers to consumers.

\subsection{Efficient Stackelberg Game-based Semantic Trading Mechanism}

Although the proposed NFT-based semantic exchange scheme can circulate semantic information between producers and consumers, it is difficult for both parties to maximize the revenue and utility of semantic information. The fixed pricing strategies may be unfair for producers when prices are lower than cost, while the unlimited pricing strategies may be disadvantaged for consumers when prices are higher than utilities. Thus, we propose a Stackelberg game-based semantic trading mechanism to provide dynamic marketplaces to trade semantic information efficiently. The Stackelberg game approach considers the realistic utility and revenue functions of producers and consumers to obtain the optimal buying and pricing strategies.

The revenue function of producers can be defined as income by providing semantic information to consumers minus their costs for computing and extracting semantic information. The income can be obtained from the required amount multiplied by the unit pricing, while the costs should consider the product of the number of CPU cycles, the CPU's frequency, the energy consumption and the network cost. 

The utility function of consumers can be given that the precision of received semantic information minuses the computing latency, and minuses the required amount multiplied by the unit pricing. The first-order derivative of the utility function should be greater than zero to incentive consumers to buy semantic information from producers.

The above functions are modeled as a Stackelberg game and analyzed the existence and uniqueness of Nash equilibrium. In the first stage, Karush-Kuhn-Tucker (KKT) conditions are utilized to help consumers determine their required amount of semantic information. The optimal solution is reported to producers for consideration. In the second stage, producers maximize their revenue given buying strategies.

\subsection{Fair Zero-Knowledge Proof-based Semantic Sharing Mechanism}

Although tokenizing semantic information can help producers to transact with consumers, it is difficult to preserve the privacy of semantic information. The reason is that blockchain is a public platform where everyone can submit and retrieve semantic information mapped in NFTs. Moreover, since semantic information can be easily duplicated, it is unfair for producers to release it before getting payments, while it is also disadvantageous for consumers to pay before verifying it. Therefore, we propose a fair and privacy-preserving zero-knowledge proof-based semantic sharing mechanism to enable fair authentic semantic sharing. Instead of directly transmitting semantic information once extracted from encoders, producers should transform or process it to protect fairness between both parties. Combining with the above mechanisms, the workflow of the zero-knowledge proof-based semantic sharing mechanism is described as follows. There are three stages corresponding to the three mechanisms. 

\begin{figure}[!t]
  \centering
  \subfigure[Communication Overhead]{\includegraphics[width=3.5in]{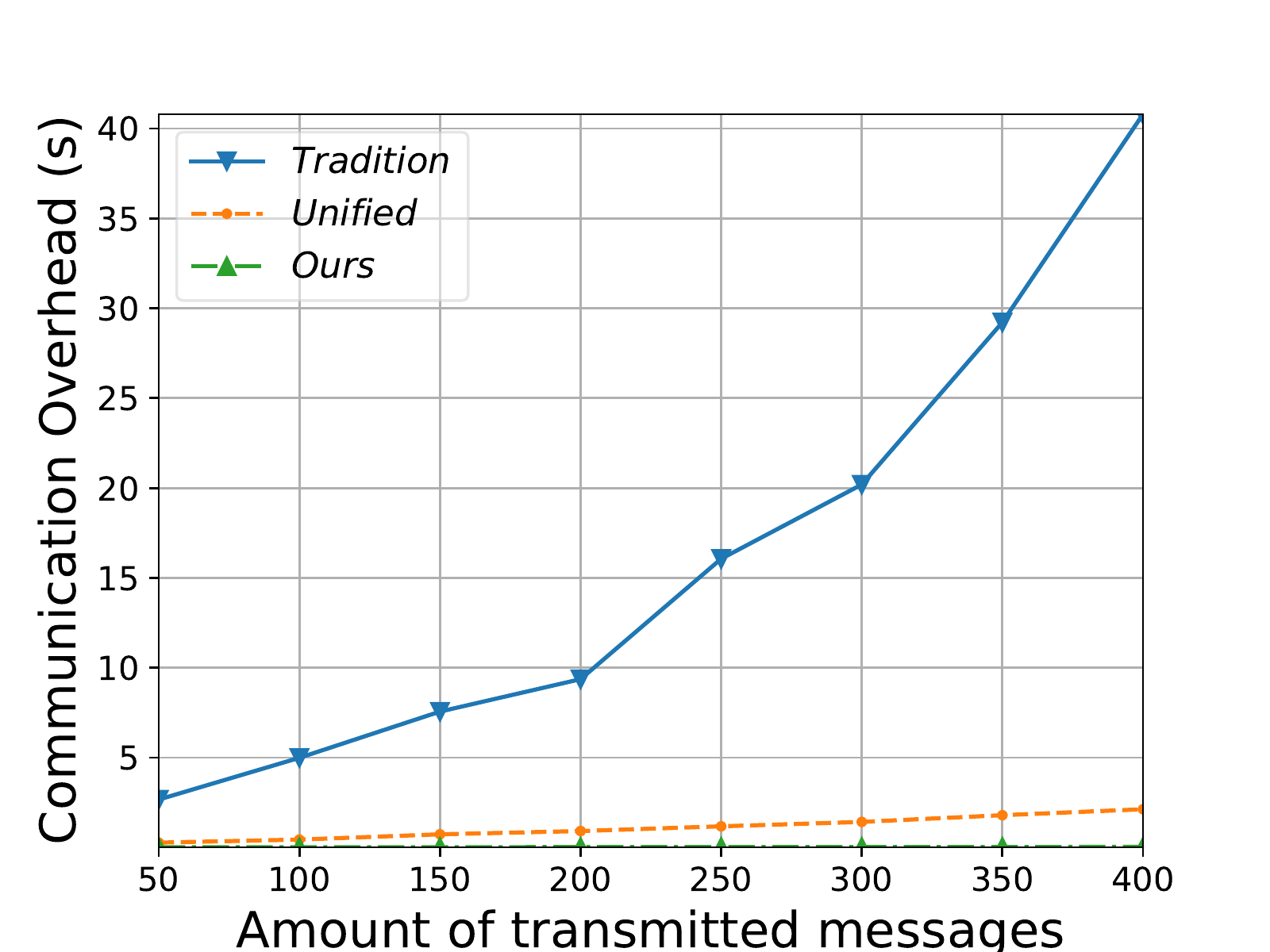}}
  \subfigure[Revenue of Semantic Trading]{\includegraphics[width=3.5in]{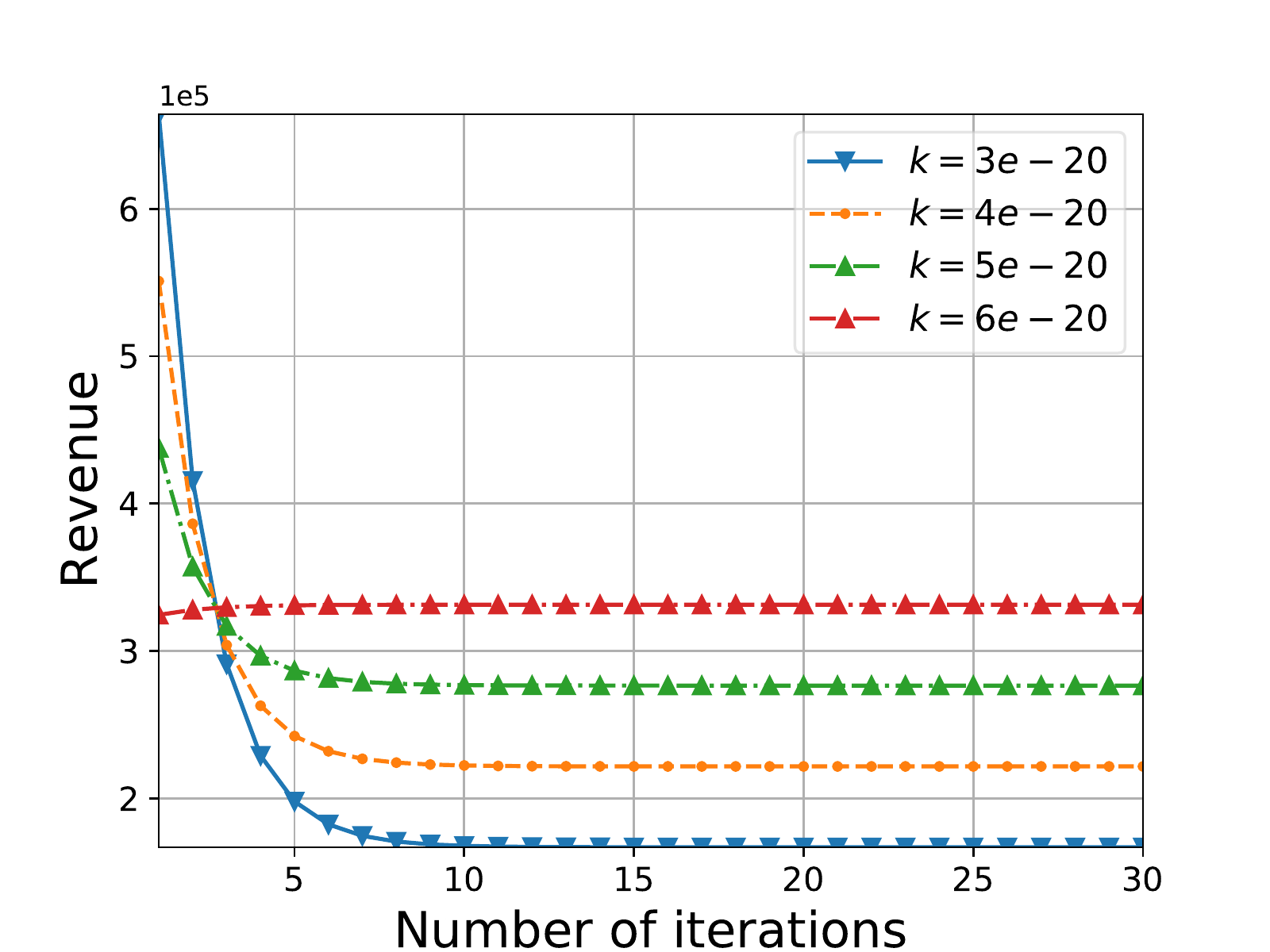}}
  \caption{An Illustration of Efficiency of The Proposed Framework}
  \label{fig_exp}
\end{figure}

\textbf{Stage \uppercase\expandafter{\romannumeral1}: Transform semantic information with zero-knowledge proof (ZKP).} Producers take a security parameter and a transform ZKP circuit as input, and output a common reference string for setup. The transform ZKP circuit carries on the logic of transformation in a private way to generate proofs of authentic information. Then they take the common reference string, and semantic information, and transformed semantic information as inputs to generate proofs.

\textbf{Stage \uppercase\expandafter{\romannumeral2}: Publishe and mint NFTs for trading given transformed semantic information.} Producers bind metadata and proofs of transformed semantic information into NFT. The transformed semantic information is not only mapped by metadata but also represents the source semantic information by proofs. 

\textbf{Stage \uppercase\expandafter{\romannumeral3}: Implement transactions between producers and consumers.} Producers and consumers configure buying and pricing strategies according to Stackelberg game-based semantic trading. After consumers pay for NFT mapped by transformed semantic information, the NFT is transferred from producers to consumers. Consumers can input proofs, transformed semantic information, and the common string to verify the authenticity of transformed semantic information without revealing the semantic information.

\section{Case Study for Urban Planning in Web 3.0}
\label{sec_case}

Urban planning activities are aided significantly by leveraging the proposed framework, which can incorporate the benefits of blockchain and semantic communication to construct Web 3.0 infrastructures. Semantic communication can reduce information overloaded to support real-time semantic interactions and minimize energy consumption for sustainable blockchain implementations. Blockchain can provide traceable semantic information status to enable semantic sharing in untrusted environments. Urban planning requires a tremendous amount of data from other counterparts to support interactions from different locations, as shown in Fig. \ref{fig_case_study}. Producers can extract the contour from images of a certain category of construction, and the corresponding snapshots (semantic information) can be transmitted to respective consumers \cite{ng2022stochastic}. Instead of transmitting original images, the extracted contour can reflect key objects and the appearance of certain urban areas to reduce information overloaded. Once urban designers obtain the contour of certain areas from other designers, they can design a panoramic for urban planning.

\begin{table*}[!t]
  \centering  
	\caption{An Overview of Open Challenges in Blockchain and Semantic Communication-Enabled Web 3.0}  
	\label{tab_challenges}  
\begin{tabular}{|p{3.8cm}|c|p{10cm}|}
\hline
  \textbf{Type} & \textbf{Layers} & \textbf{Open Problem} \\
\hline
\hline
\multirow{3}{*}{\makecell{Blockchain-Enabled \\ Web 3.0}} 
  & On-Chain & Storage Structure Design, Consensus Protocol Design, Regulation Solutions, Quantum-Resistance Protocol Design\\
  \cline{2-3}
  & Off-Chain & Cross-Chain Solutions, Oracle Protocol Design, Rollup Protocol Design, Zero-Knowledge Proof Layer 2 Solutions\\ 
   \cline{2-3}
  &  Collaboration  & Decentralized Identity, Decentralized Finance, Metaverse, Decentralized Autonomous Organization\\
   \cline{2-3}
\hline
\multirow{3}{*}{\makecell{Semantic Communication\\-Enabled  Web 3.0}} &   Theory & Explainability, Impact of Transmission Rate, Capacity of Semantic Channel \cite{shi2021semantic}, Quantity of Semantic Information \cite{luo2022semantic} \\
 \cline{2-3}
  &   Metric  & Importance of Semantic Information, Tradeoff Between Accuracy and Communication Overhead \cite{yang2022semantic}, Inconsistent Background Knowledge Between Source and Destination \cite{luo2022semantic} \\ 
   \cline{2-3}
  &  Application  & Implementation of Semantic Communication, Multi-User Collaboration, Video Transmission, Personalization  \\
   \cline{2-3}
\hline
\multirow{3}{*}{\makecell{Blockchain and Semantic\\-Enabled Web 3.0}} & Architecture & Unified, Secure, Efficient, Decentralized, Fault-Tolerant, Verifiable, Interoperable, Regulated \\
 \cline{2-3}
&   Collaboration  & Verification and Price Bubbles of Semantic Information, Semantic Interoperability, Task-Oriented Layer 2 Solutions \\ 
 \cline{2-3}
& Application & Metaverse, Digital Twin, Health Services, Digital Asset Management \cite{lin2022unified}, Automotive Industry \\
 \cline{2-3}
\hline
\hline
\end{tabular}
\end{table*}

However, information circulated in the urban planning activities can be duplicated easily, which makes that other urban planning designers are not willing to exchange it without any benefit and profit. For the same reason, it is unfair for producers to release semantic information first, which makes other urban planning designers can pay nothing to receive the core contour of certain areas. Moreover, since semantic information is valuable, it is necessary to construct a trading mechanism to guide semantic exchange in urban planning activities. 

The proposed framework is based on the above two infrastructures to build fair and efficient semantic exchange mechanisms, as shown in Fig. \ref{fig_case_study}. The designed zero-knowledge proof-based semantic sharing mechanism can implement fair and efficient transactions among urban designers by downsizing semantic information to exchange the contour of semantic information without releasing core designs first. It utilizes the NFT-based semantic exchange scheme to facilitate downsized semantic information produced from spatial data to circulation in urban planning activities. The proposed Stackelberg game-based semantic trading mechanism can facilitate urban planning activities constructing efficient trading marketplaces to facilitate the circulation of semantic information.




To evaluate the proposed framework and mechanisms, we implement the simulation case to illustrate the efficiency, as exhibited in Fig. \ref{fig_exp}. Fig. \ref{fig_exp}(a) shows the communication overhead among tradition communication (Tradition), blockchain-based semantic communication (Unified) \cite{lin2022unified}, and our proposed mechanism (Ours). The blockchain module is supported by a practical byzantine fault tolerance algorithm with four nodes. The semantic module is implemented by Pytorch. The size of the transmitted messages among those mechanisms is 1 MB, 100 KB, and 1 KB, referring to Fig. \ref{fig_case_study}. The amount of semantic information represents exchanged number of messages. The communication overhead is the time that semantic information is recorded in the blockchain module. As shown in Fig. \ref{fig_exp}(a), the proposed mechanism is more efficient than the compared mechanisms. Fig. \ref{fig_exp}(b) illustrates how the revenue of producers is affected by different parameters of the semantic trading mechanism as the number of iterations increases. With the increase of parameters, the costs take up more weight in the revenue function, which causes a decrease in the income of producers.

\section{Open Challenges and Future Directions}
\label{sec_future}

We overview recent open challenges in terms of blockchain, semantic communication, and Web 3.0, as shown in Table \ref{tab_challenges}. Then we also elaborate on several of them in the following.

\textbf{Verification of Semantic Information:} In scenarios that require high privacy and security, e.g.,  automotive industries, verification of semantic information is critical to perform operations correctly. It is impossible to publish semantic information on public Web 3.0 platforms, and leverage semantic information without any verification. Thus, zero-knowledge proof combined with pragmatic functions of semantic communication is expected to share private and authentic semantic information to protect privacy and security in Web 3.0. Besides, since proofs generated by ZKP can be verified given mathematical relationships, reputation mechanisms can be utilized to classify and exclude malicious Web 3.0 participants. 

\textbf{Price Bubbles of NFT-mapped Semantic Information:} Although semantic information can be considered as NFT-based Web 3.0 assets circulating in the marketplaces, it is difficult to avoid price bubbles in current market mechanisms, like OpenSea \cite{white2022characterizing}. The English Dutch auction mechanism \cite{deck2020designing} is promising to eliminate price bubbles in the circulation of semantic information.

\textbf{Semantic Interoperability:} Since Web 3.0 is integrated with blockchain and semantic communication, semantic interoperability not only includes semantic information circulated among multiple blockchains but also interacted with multiple semantic tasks. The former can be implemented by cross-chain technologies combined with verification, while the latter should consider how to extract the same characteristics between different types of semantic tasks.

\textbf{Task-Oriented Layer 2 Solutions:} Participants have to continuously exchange semantic information after perceiving environments to improve communication quality, which is time and resource-consuming. Current Layer 2 solutions may be a practical method to improve the performance of blockchain and semantic communication-based Web 3.0, which can provide off-chain interactions in a decentralized and traceable way.

\section{Conclusion}
\label{sec_conclusion}

In this paper, we present our understanding and research gaps in Web 3.0, Blockchain, and Semantic Communication. To this end, we propose a fair and efficient blockchain-based semantic exchange framework for the participatory economy. In this framework, we utilize NFT to circulate semantic information in the blockchain to construct value networks. After that, we consider optimal buying and pricing strategies for counterparts transacting with semantic information by a Stackelberg game approach. We also take fairness into account and leverage the zero-knowledge proof to enable authentic semantic sharing. A case study of urban planning and simulation results are implemented to illustrate the efficiency of the proposed framework. Finally, key challenges, opportunities, and future research related to blockchain and semantic communication for Web 3.0 are discussed.

\bibliographystyle{IEEEtran}
\bibliography{ref}

\noindent{\text{BIOGRAPHIES}}

{\small \noindent {\textbf{Yijing Lin}} is currently pursuing the Ph.D degree at the State Key Laboratory of Networking and Switching Technology, Beijing University of Posts and Telecommunications, China (e-mail: yjlin@bupt.edu.cn).}

\vspace{0.1cm}
{\small \noindent {\textbf{Zhipeng Gao}} is the corresponding author of this paper. He is a professor with the State Key Laboratory of Networking and Switching Technology, Beijing University of Posts and Telecommunications, China (e-mail: gaozhipeng@bupt.edu.cn). }

\vspace{0.1cm}
{\small \noindent {\textbf{Yaofeng Tu}} is the Director of Center Institute of ZTE, China (e-mail: tu.yaofeng@zte.com.cn). }

\vspace{0.1cm}
{\small \noindent {\textbf{Hongyang Du}} is currently pursuing the Ph.D degree at the School of Computer Science and Engineering, Nanyang Technological University, Singapore (e-mail: hongyang001@e.ntu.edu.sg).}

\vspace{0.1cm}
{\small \noindent {\textbf{Dusit Niyato}} is a professor with the School of Computer Science and Engineering, Nanyang Technological University, Singapore (e-mail: dniyato@ntu.edu.sg).}

\vspace{0.1cm}
{\small \noindent {\textbf{Jiawen Kang}} is a professor with the School of Automation, Guangdong University of Technology, China (e-mail: kjwx886@163.com).}

\vspace{0.1cm}
{\small \noindent {\textbf{Hui Yang}} is a professor with the State Key Laboratory of information Photonics and Optical Communications, Beijing University of Posts and Telecommunications, China (e-mail: yanghui@bupt.edu.cn).}

\end{document}